\documentclass[12pt]{iopart}

\usepackage{graphicx}  
\begin{document}

\title{Stable pulse generation in bias pumped gain-switched fiber laser}

\author{Fuyong Wang}

\address{School of Information and Electrical Engineering, Hebei University of Engineering, Handan 056038, China}
\ead{jiaoyi@sjtu.edu.cn}
\vspace{10pt}

\begin{abstract}
We demonstrate gain-switched fiber laser with bias pumping plays an important role in regulating chaotic relaxation spikes. Under certain conditions the profile of output pulse from gain-switched fiber laser can keep the same shape as that of pump pulse. Bias pumping technique may have significance in increasing output stable pulse energy in gain-switched fiber laser. Gain-switched fiber laser with bias pumping may be interesting for some applications in micro-processing and thin film removal.
\end{abstract}

\vspace{2pc}

\section{Introduction}

Pulsed laser operation have received a significant attention in recent years for many applications in medicine and manufacturing. Typical methods to produce laser pulses are Q-switching, cavity dumping, passive and active mode-locking. Gain-switching of fiber lasers has been established as an alternative pulsing method due to the simplicity and all-fiber integration. There is no need for additional active optical element and no need for free-space coupling to an acousto-optical modulator. It can be a very simple and robust technique to produce laser pulses. Together with fiber technology it can produce compact and reliable lasers appropriate for industrial applications such as micro-processing. Compared to Q-switched fiber lasers, gain-switched fiber lasers have their own distinct advantages including the simplicity in operation and extreme flexibility in pulse repetition rate tuning. Gain-switched fiber laser gets an increasing attention due to the decreasing cost of pump diodes and the all-fiber construction \cite{Vid2014,intro1,intro2,introt,one1,one2}.

Gain-switched fiber laser gets rapidly improved with the development of high-power pump diodes and fiber laser technology \cite{Maryashin2006,Larsen2014}. However, there are still some problems in gain-switched fiber laser, one of which is chaotic relaxation spike phenomenon. Gain-switched fiber laser has been investigated by a number of researchers but the output is mostly chaotic \cite{one4,one5}. Some numerical studies have been carried out to derive the origin of the trailing spike phenomenon and pump conditions have been summarized to avoid the pulse shape distortion \cite{intro3}. They pointed out longer pump pulses could induce more trailing spikes. In order to regulate the conventionally chaotic spiking in gain-switched fiber laser and obtain a stable pulse train, fast gain-switching and resonant pumping are required \cite{one3,Swiderski2013}. For typical three or four level laser systems, however, fast gain-switching is deterred by the relaxation process from the pump absorption energy level to the emission energy level. Fast gain-switching is therefore difficult, and relaxation oscillation between the photon density and the excitation population leads to chaotic pulsation.

In conventional gain-switched fiber laser, in order to achieve stable output laser pulse (or avoid chaotic relaxation spike phenomenon) pump pulse peak power and duartion are usually low and short. In this letter, we explore whether stable output pulse can be generated in gain-switched fiber laser without shortening pump pulse duration or decreasing pump power. With numerical simulation we show chaotic relaxation spike in gain-switched fiber laser caused by long pump duration and high peak power of pump. Then we propose a bias pumped gain-switched fiber laser to regulate these spikes. 

\section{Typical configuration and rate equations of the gain-switched fiber laser}

\begin{figure*}[htbp]
\centerline{\includegraphics[width=5in]{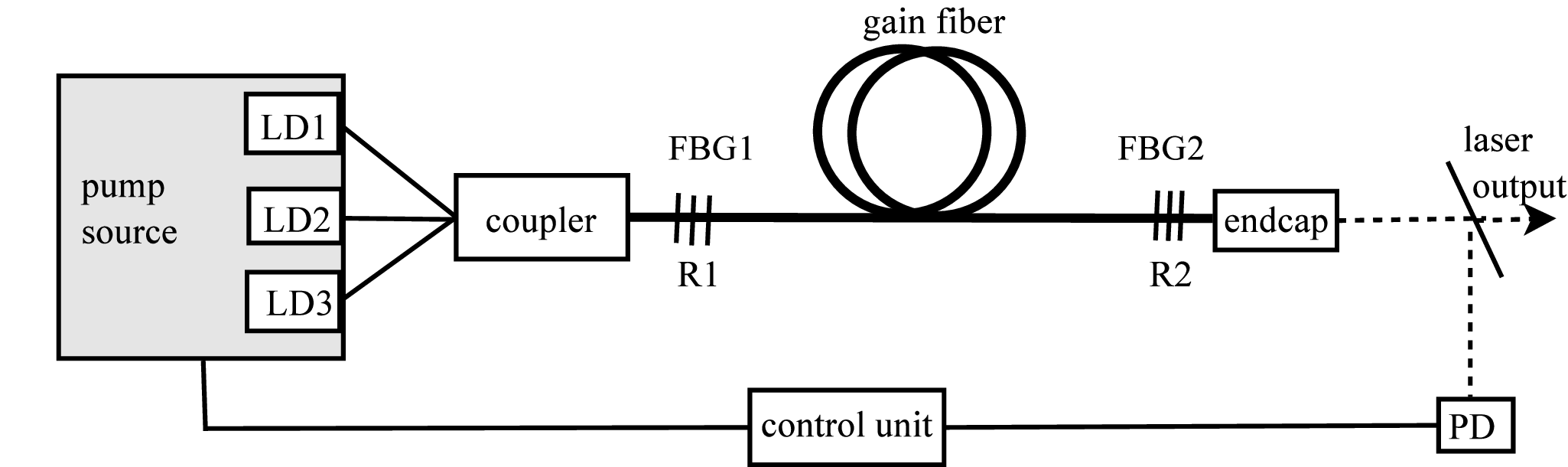}}
\caption{Schematic representation of a gain-switched fiber laser.}\label{setup}
\end{figure*}

The gain-switched fiber laser, which are used in our simulation, are sketched on figure \ref{setup}. To simplify the handling and enhance the mechanical stability, it is appropriate to use fiber bragg gratings (FBG) instead of external mirrors and fiber coupled diode lasers to pump the fiber.

The laser is powered by several laser diodes (LD) working at a desired wavelength. In principle the pumping power can be easily increased by adding further pumping units. A fraction of the output laser is deflected onto a photodiode (PD) to provide a signal for the control unit which is composed of a logic circuit, a trigger and a pulse generator. The pulse generator switches on the pumping diodes at the desired repetition rate. The laser pulse causes a trigger signal in the logic circuit that immediately switches off the pumping source. By adjusting the repetition rate of the pulse generator, the pumping starts again and cycle repeats. The electronics consist of a high-power laser diode driver capable of driving the laser diode array in continuous-wave (CW) and pulsed operation. 

In the following simulation, we use a double-cladding highly doped ytterbium fiber as gain medium with a length of 4 m. To pump the Yb-doped fiber, we use pulsed diode lasers with a wavelength of about $\lambda _p\approx$980 nm. The reflectivity of the fiber bragg grating 1 (R1) is about 99 $\%$ and of the grating 2 (R2) is about 10 $\%$ at the laser emission wavelength of $\lambda _s\approx$ 1090 nm.

The time- and space-dependent rate equations for the laser system shown in Fig. \ref{setup} are based upon the traveling wave model and given by \cite{two1}
\begin{equation}
N=N_1+N_2,\label{e1}
\end{equation}
\begin{equation}
\frac{\partial N_2}{\partial t}+\frac{N_2}{\tau }=\frac{\Gamma _p\lambda _p}{hcA}[\sigma _{ap}N_1-\sigma _{ep}N_2]P_p+\frac{\Gamma _s\lambda _s}{hcA}[\sigma _{as}N_1-\sigma _{es}N_2](P_{sf}+P_{sb}),\label{e2}
\end{equation}
\begin{equation}
\frac{\partial P_p}{\partial z}+\frac{1}{\upsilon _p}\frac{\partial P_p}{\partial t}=\Gamma _p[\sigma _{ep}N_2-\sigma _{ap}N_1]P_p-\alpha _pP_p,
\end{equation}
\begin{equation}
\frac{\partial P_{sf}}{\partial z}+\frac{1}{\upsilon _s}\frac{\partial P_{sf}}{\partial t}=\Gamma _s[\sigma _{es}N_2-\sigma _{as}N_1]P_{sf}-\alpha _sP_{sf}+2\sigma _{es}N_2\frac{hc^2}{\lambda _{s}^3}\Delta \lambda _s, 
\end{equation}
\begin{equation}
-\frac{\partial P_{sb}}{\partial z}+\frac{1}{\upsilon _s}\frac{\partial P_{sb}}{\partial t}=\Gamma _s[\sigma _{es}N_2-\sigma _{as}N_1]P_{sb}-\alpha _sP_{sb}+2\sigma _{es}N_2\frac{hc^2}{\lambda _{s}^3}\Delta \lambda _s,\label{e5}
\end{equation}
where N is the total doping concentration. N$_1$ and N$_2$ are the lower and upper population concentrations, respectively. $\tau $
is the fluorescence lifetime. $\Gamma _p$ and $\Gamma _s$ are the overlap factors between the pump and signal and the doped fiber area, respectively. $\lambda _p$ and $\lambda _s$ are the pump and signal free-space wavelengths, respectively. h is the Planck constant. c is the speed of light in vacuum. A is the core area. $\sigma _{ap}$ and $\sigma _{ep}$ are the absorption and emission cross sections of the pump power, respectively. $\sigma _{as}$ and $\sigma _{es}$ are the absorption and emission cross sections of the signal powers, respectively. P$_p$ is the pump power. P$_{sf}$ and P$_{sb}$ are the forward and backward signal powers, respectively. $\upsilon _p$ and $\upsilon _s$ are group velocity of the pump and the laser pulses propagating in fiber. $\alpha _s$ and $\alpha _p$ are the attenuation of the signal and pump powers, respectively, and $\Delta \lambda _s$ is the bandwidth of the amplified spontaneous emission (ASE) at around 1 $\mu$m.

The above equations (\ref{e1})-(\ref{e5}) are suitable for solving both CW and pulsed Yb-doped fiber lasers. These equations are governed by the boundary conditions, which physically represent the feedback provided by both ends of the laser cavity. The boundary  conditions for equations (\ref{e1})-(\ref{e5}) are given by
\begin{equation}
P_p(z=0,t)=W_0,
\end{equation}
\begin{equation}
P_{sf}(z=0,t)=R_1P_{sb}(z=0,t),
\end{equation}
\begin{equation}
P_{sb}(z=L,t)=R_2P_{sf}(z=L,t),
\end{equation}
\begin{equation}
P_{out}(z=L,t)=(1-R_2)P_{sf}(z=L,t),
\end{equation}
where R$_1$ and R$_2$ are the reflectivities of FBG 1 and FBG 2, respectively, and P$_{out}$ is the output signal power.

The main parameters used in the simulations are summarized in the Table \ref{t1}. 
\begin{table}[htbp]
\newcommand{\tabincell}[2]{\begin{tabular}{@{}#1@{}}#2\end{tabular}}
  \caption{\label{t1}Main parameters used in the simulations \cite{two1}}
  \begin{center}
    \begin{tabular}{ccccccccc}
    \hline
     \tabincell{c}{Parameter}& Value& Parameter& Value  \\
    \hline
    $\lambda _p$ &980 nm &$\sigma _{ap}$ &2.5$\times 10^{-24}$ m$^2$ \\
    $\lambda _s$ & 1090 nm&$\alpha _s$ & 5$\times 10^{-3}$ m$^{-1}$ \\
    N & 4$\times 10^{26}$ m$^{-3}$&$\alpha _p$ & 0.39 m$^{-1}$ \\
    $\tau$  & 1 ms&$\Delta \lambda _s$ & 20 nm\\
    A &2.83$\times 10^{-11}$ m$^2$& $\Gamma _s$  & 0.75\\
    $\sigma _{es}$ & 3.5$\times 10^{-25}$ m$^2$&$\Gamma _p$ & 0.0023\\
    $\sigma _{as}$ &2.0$\times 10^{-27}$ m$^2$& $R_1$ & 0.99\\
    $\sigma _{ep}$ & 3.0$\times 10^{-24}$ $m^2$&$R_2$  & 0.1 \\
    \hline
    \end{tabular}
  \end{center}
\end{table}

\section{Simulations and discussion on pulses characteristics}

\begin{figure*}[htbp]
\begin{minipage}{0.5\linewidth}
\centerline{\includegraphics[width=7cm,height=5cm]{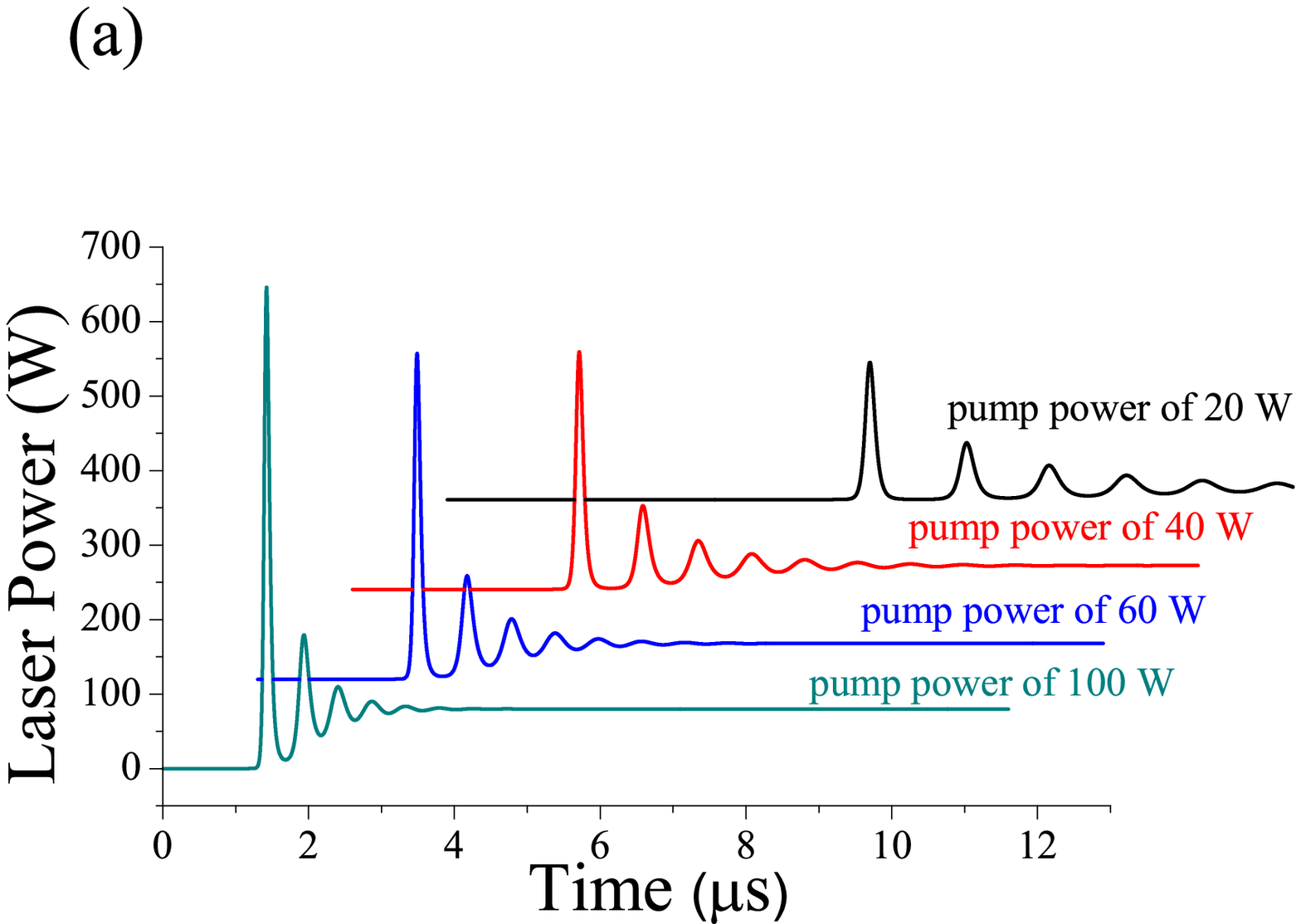}}
\centerline{ }
\end{minipage}
\hspace{-0.35in}
\begin{minipage}{0.5\linewidth}
\centerline{\includegraphics[width=7cm,height=5cm]{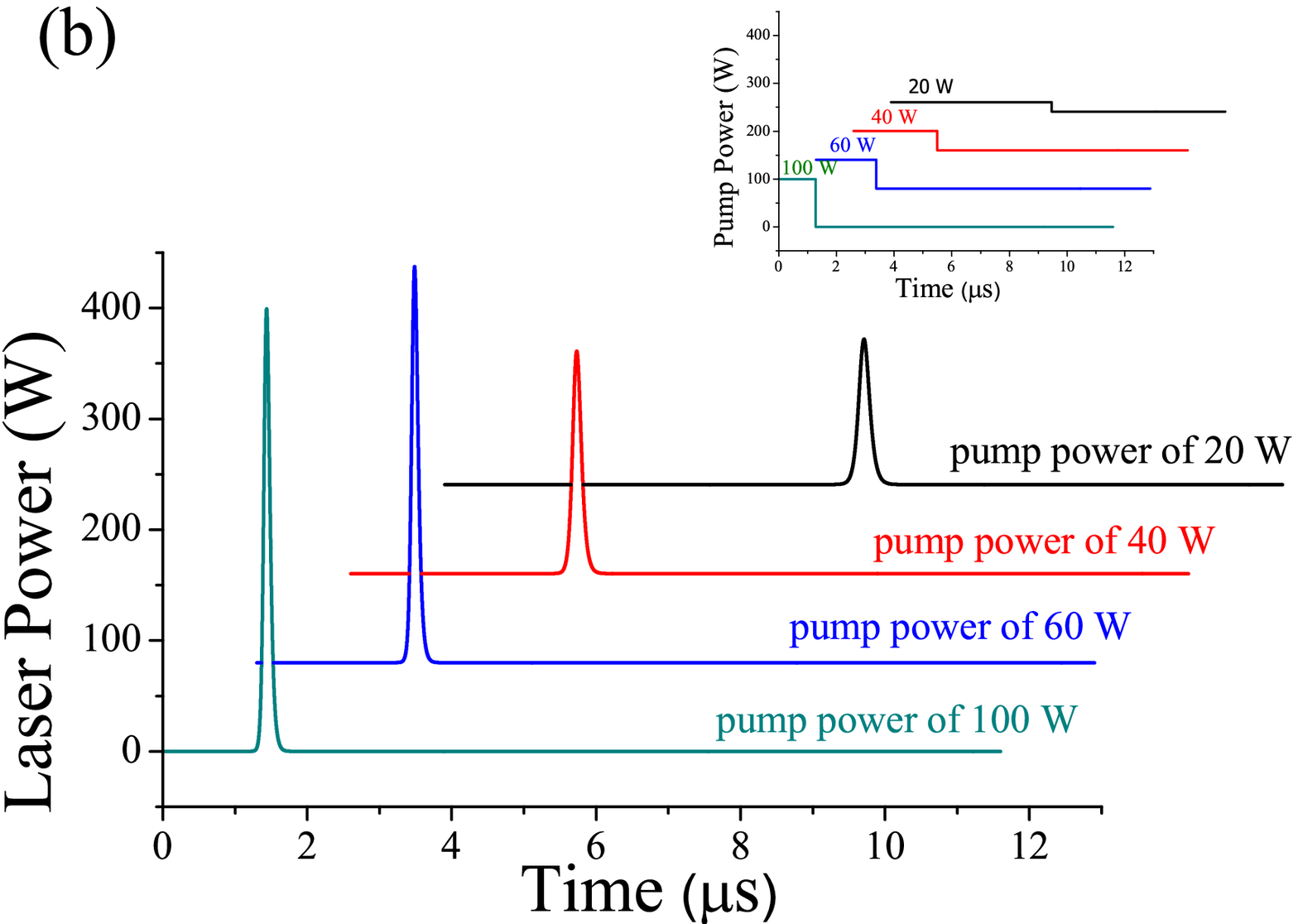}}
\centerline{ }
\end{minipage}
\vfill
\begin{minipage}{0.5\linewidth}
\centerline{\includegraphics[width=7cm,height=5cm]{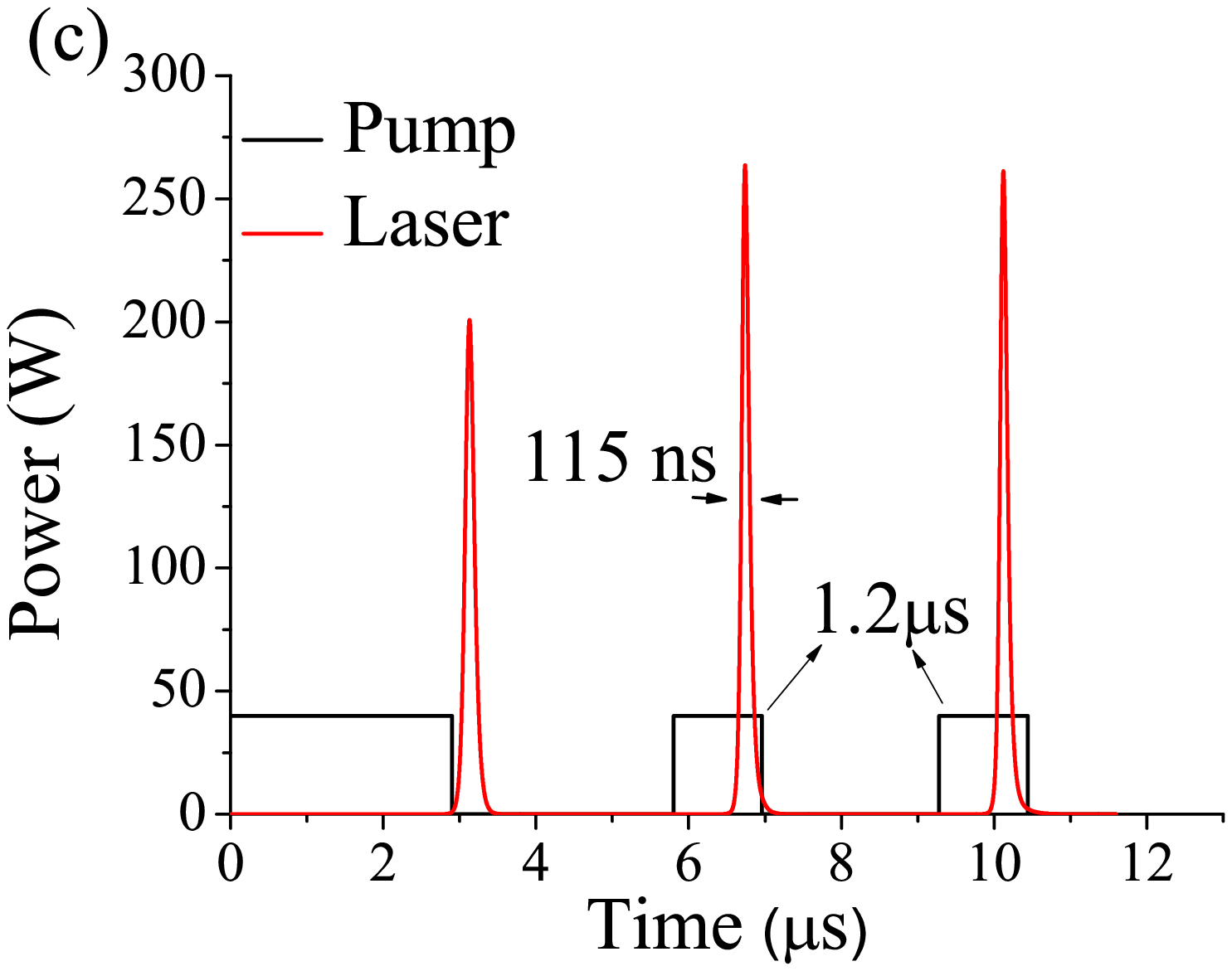}}
\centerline{}
\end{minipage}
\hspace{-0.35in}
\begin{minipage}{0.5\linewidth}
\centerline{\includegraphics[width=7cm,height=5cm]{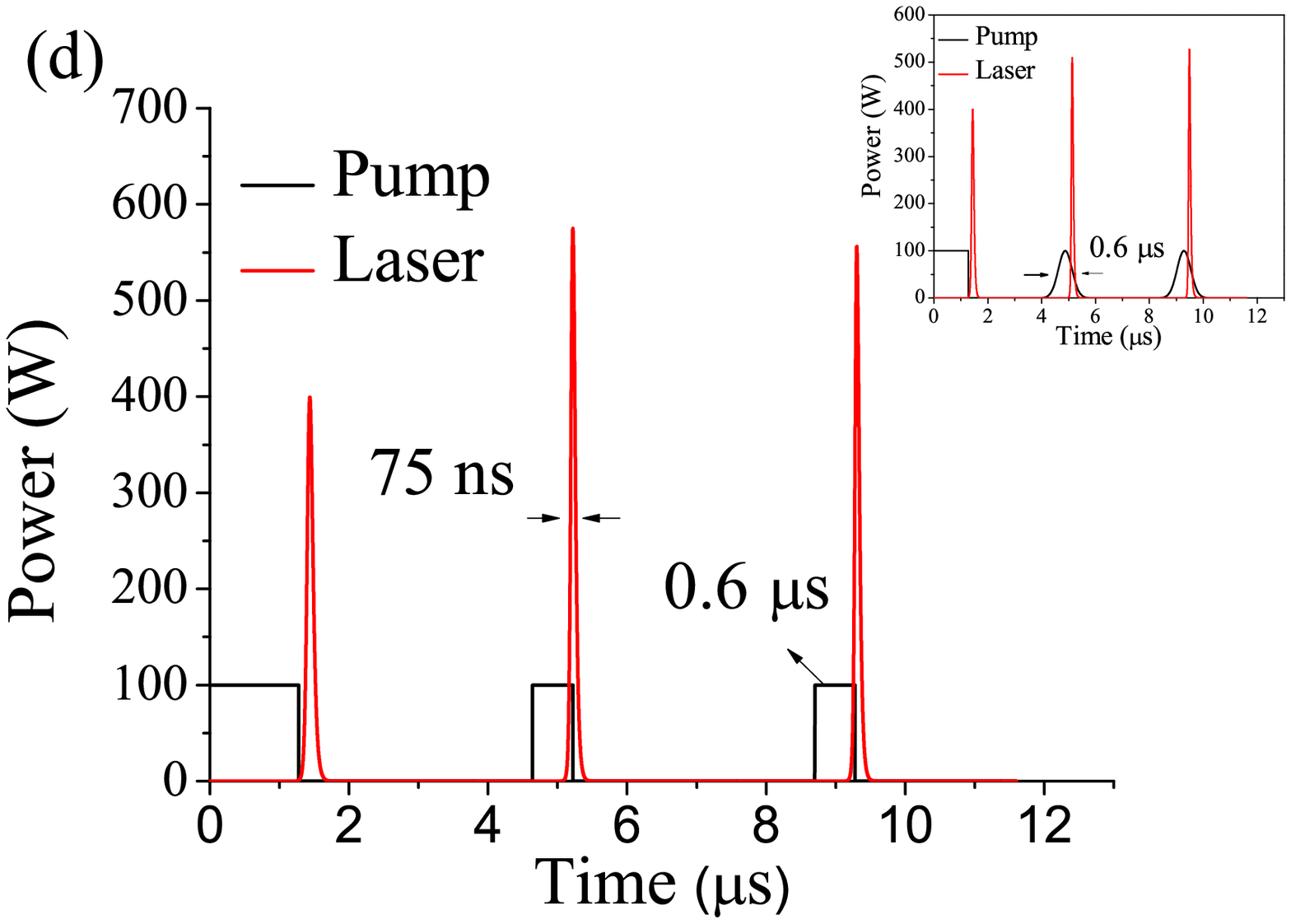}}
\centerline{ }
\end{minipage}
\caption{Temporal chracteristics of the outputs (a) at different CW operation pump powers of 20 W, 40 W, 60 W and 100 W; (b) at different pump powers of 20 W, 40 W, 60 W and 100 W with the corresponding pumping time lasting 5.5 $\mu$s, 2.9 $\mu$s, 2.1 $\mu$s and 1.3 $\mu$s. Temporal chracteristics of the output pulse train (c) with pump peak power of 40 W and duration of 1.2 $\mu$s; (d) with pump peak power of 100 W and duration of 0.6 $\mu$s.}
\label{sp}
\end{figure*}

We first investigate the temporal characteristics of output laser under different CW pump powers and the results are shown in figure \ref{sp}(a). As long as pump power is above threshold the final state of the output laser is CW steady state. Before getting steady state, all of the output lasers experience relaxation oscillation as shown in figure \ref{sp}(a). The first relaxation oscillation spike occurs at 1.4 $\mu$s with pump power of 100 W (2.2 $\mu$s with pump power of 60 W, 3.1 $\mu$s with pump power of 40 W and 5.8 $\mu$s with pump power of 20 W). The higher of pump power, the earlier of the relaxation oscillation appears. If pump is switched off at an appropriate time, other relaxation oscillation spikes behind the first one can be supressed and a stable pulse is obtained as can be seen from figure \ref{sp}(b). The output pulses of figure \ref{sp}(b) are corresponding to the first spikes of the relaxation oscillations of figure \ref{sp}(a), respectively. This leads to another understanding of gain-switching as the first relaxation oscillation pulse of the CW operation. 

When the pump is switched on and off periodically at a certain repetition, a sequence of gain-switched pulses are generated as shown in figure \ref{sp}(c) and figure \ref{sp}(d). In order to get stable gain-switched pulse, pump pulse duration is about 1.2 $\mu$s (0.6 $\mu$s) with pump peak power of 40 W (100 W) as shown in figure \ref{sp}(c) (figure \ref{sp}(d)). When the pump pulse is gaussion shape with duration of 0.6 $\mu$s, stable output pulse can still be generated with pump peak power of 100 W (inset of figure \ref{sp}(d)). The simulation results clearly demonstrate that a short pump duration is required under high pump peak power for stable pulse outputs. 

\begin{figure*}[htbp]
\begin{minipage}{0.5\linewidth}
\centerline{\includegraphics[width=7cm,height=5cm]{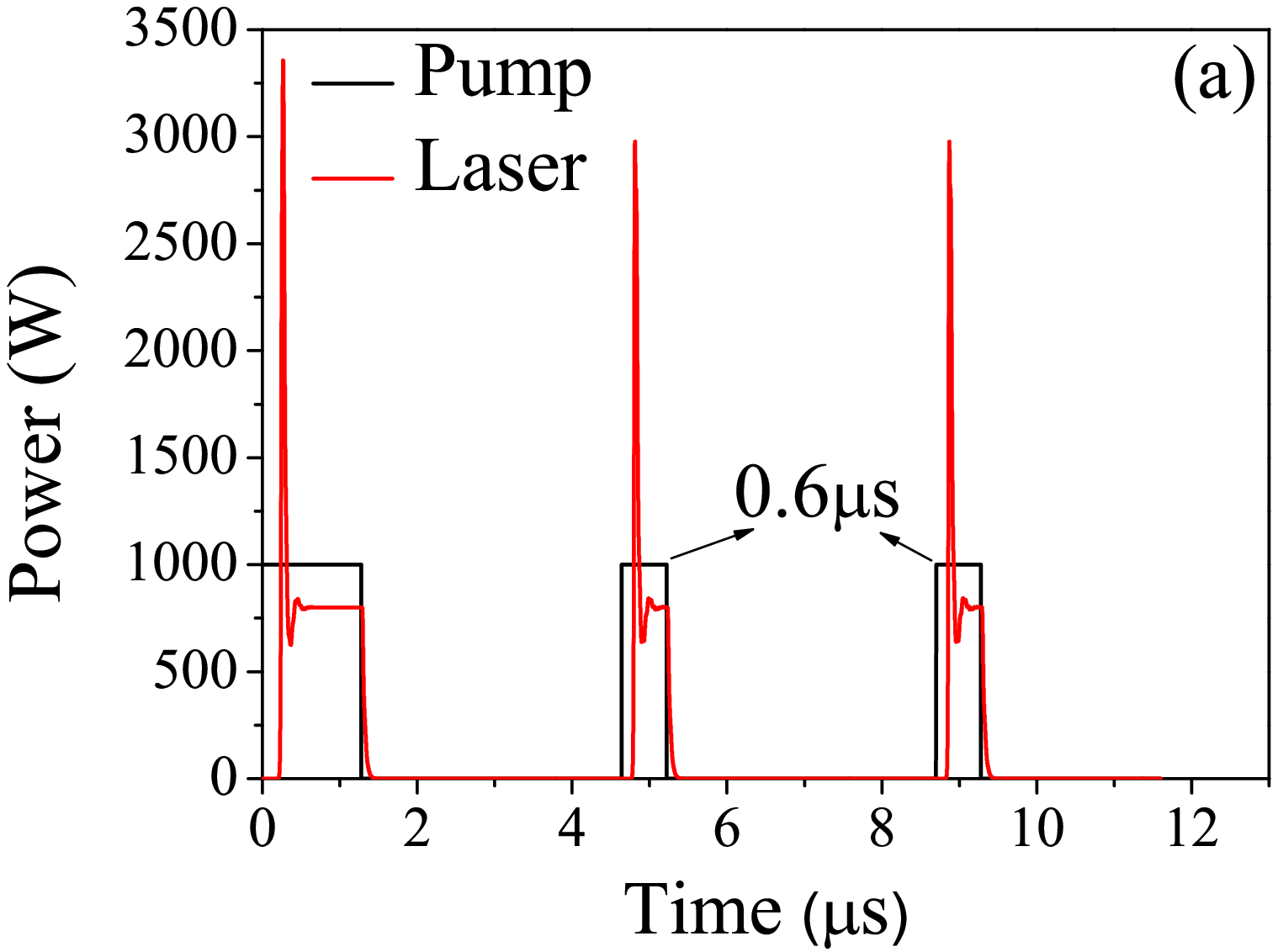}}
\centerline{ }
\end{minipage}
\hspace{-0.35in}
\begin{minipage}{0.5\linewidth}
\centerline{\includegraphics[width=7cm,height=5cm]{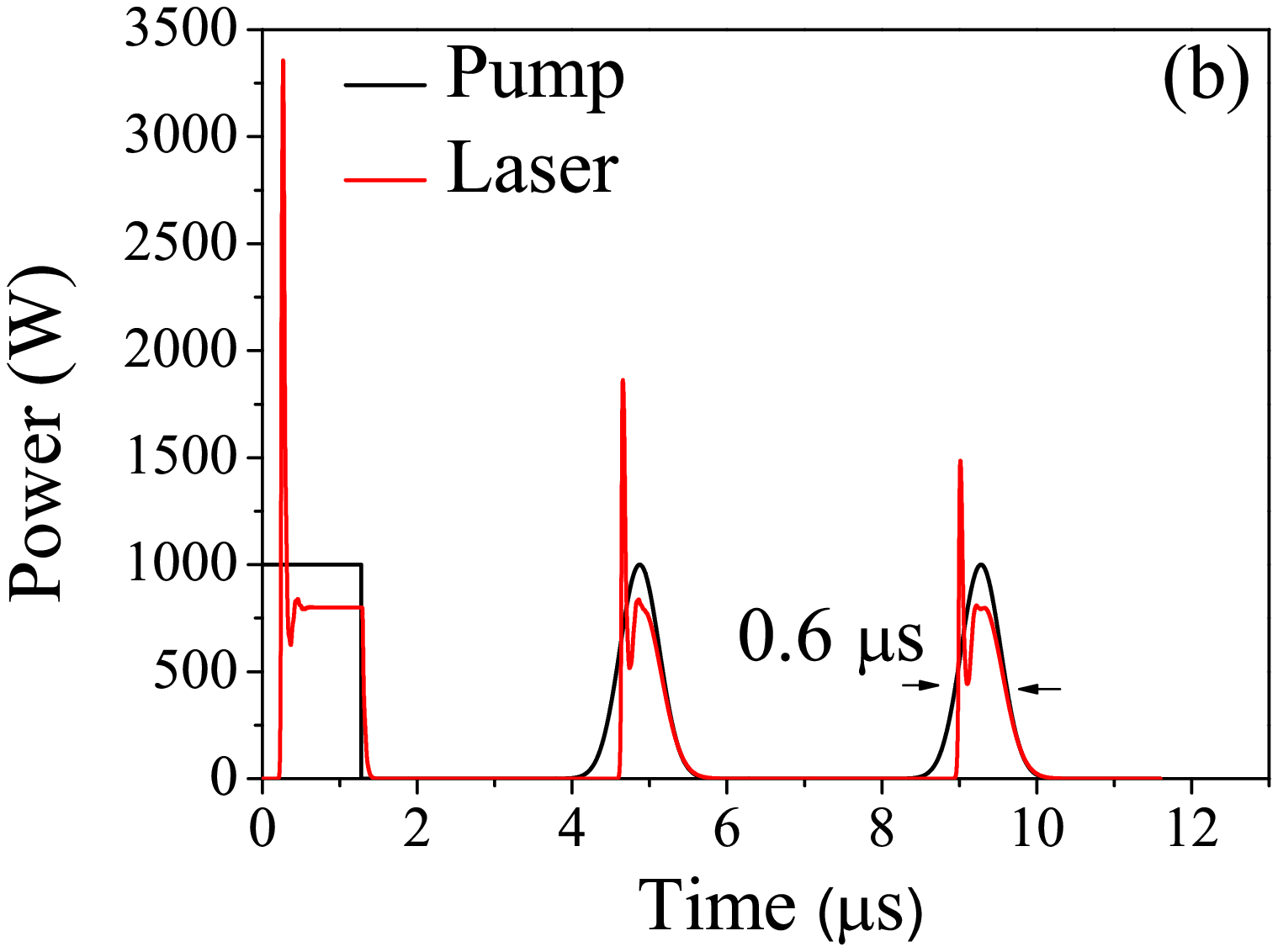}}
\centerline{ }
\end{minipage}
\vfill
\begin{minipage}{0.5\linewidth}
\centerline{\includegraphics[width=7cm,height=5cm]{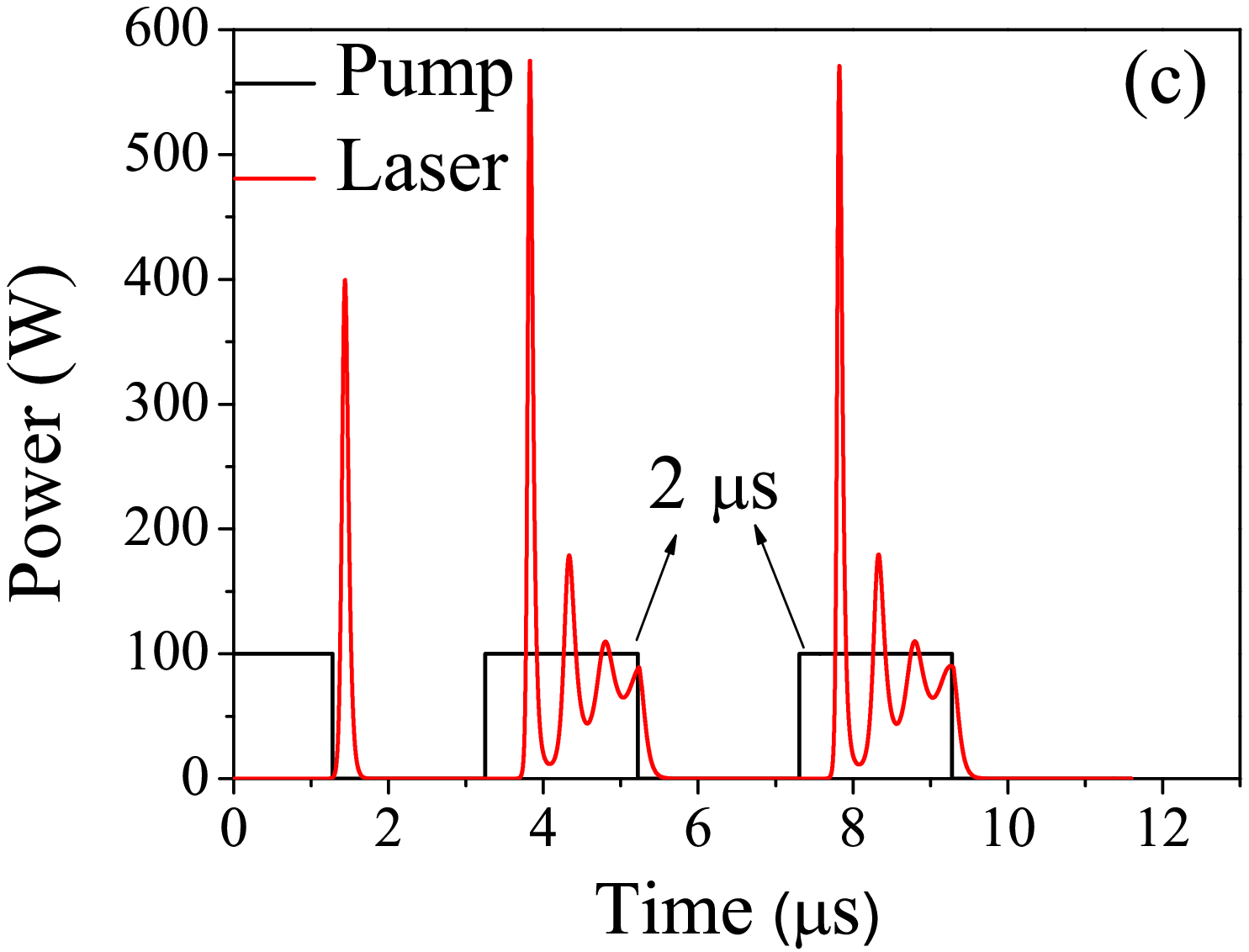}}
\centerline{}
\end{minipage}
\hspace{-0.35in}
\begin{minipage}{0.5\linewidth}
\centerline{\includegraphics[width=7cm,height=5cm]{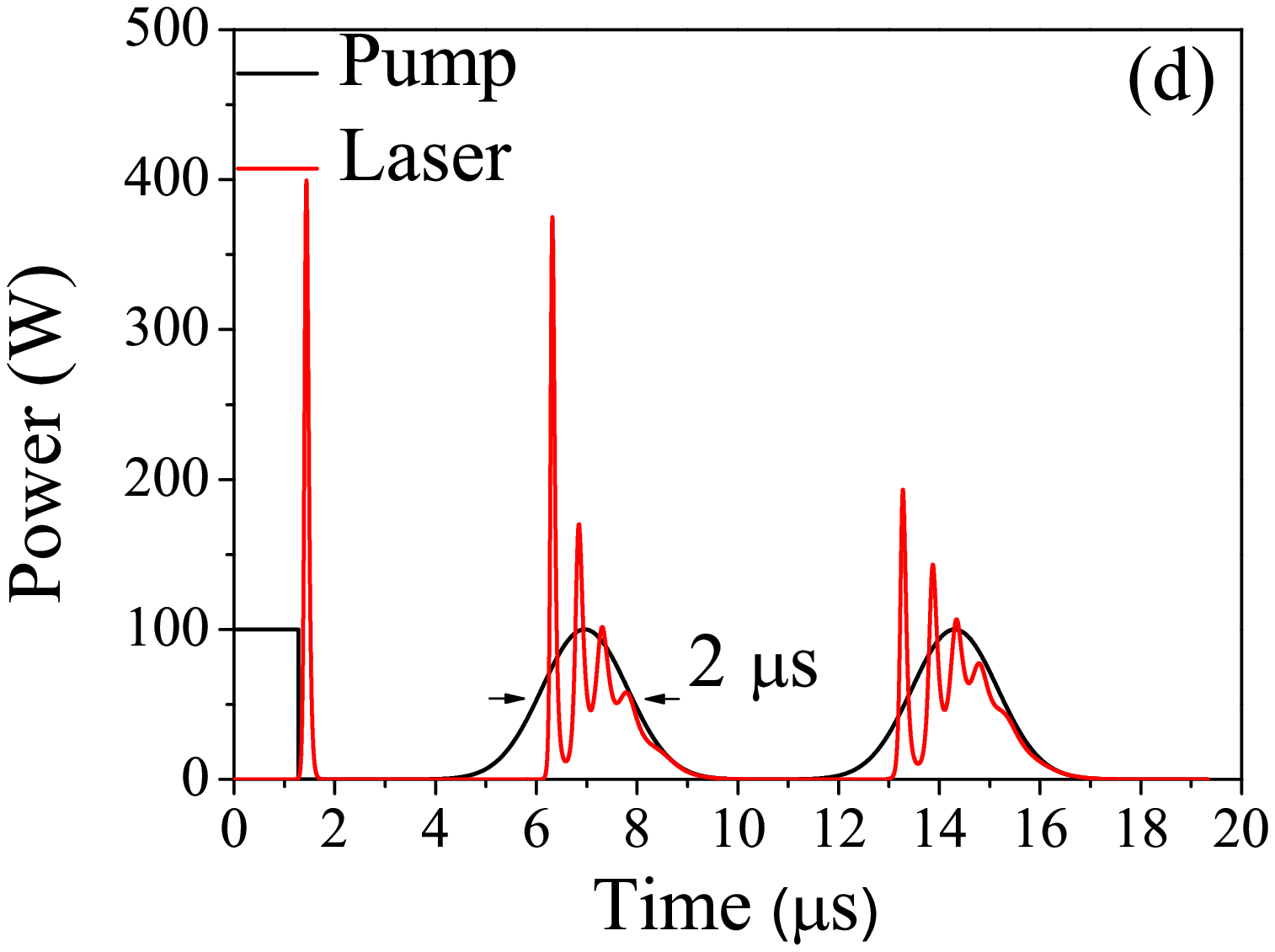}}
\centerline{ }
\end{minipage}
\caption{Temporal chracteristics of the outputs under pump power of 1000 W and duration of 0.6 $\mu$s (a) with rectangular pump pulse shape; (b) with gaussion pump pulse shape. Temporal chracteristics of the outputs under pump power of 100 W and duration of 2 $\mu$s (c) with rectangular pump pulse shape; (d) with gaussion pump pulse shape.}
\label{sr}
\end{figure*}

If we increase the peak power of pump to 1000 W and keep its duration the same as that of in figure \ref{sp}(d), the output stable pulse in figure \ref{sp}(d) becomes "figure-of-h" pulse shape in figure \ref{sr}(a). With pump peak power increasing, the signal pulse becomes unstable and breaks into multipulsing, as the excitation population becomes too large to be depleted by the first signal pulse. Keeping peak power of 1000 W unchanged and just changing rectangular pump profile to gaussian pump profile, the output gain-switched laser is still "figure-of-h" profile as shown in figure \ref{sr}(b). The change of stable gain-switched pulse to chaotic "figure-of-h" pulse is directly related to the increasing of pump power and independent of pump pulse shape. One way to obtain stable pulse under 1000 W peak power pumping is decreasing the pump pulse duration.

If we enlarge the duration of pump to 2 $\mu$s and keep its peak power the same as that of in figure \ref{sp}(d), the output stable pulse in figure \ref{sp}(d) becomes unstable and several spikes appear as shown in figure \ref{sr}(c) and figure \ref{sr}(d). Regardless of rectangular pump profile or gaussion pump profile, long pump duration can lead to chaotic relaxation spike phenomenon. 

According to the investigation presented above, high peak power pumping and long pump pulse duration lead to chaotic pulsation. High pump power tends to result in rapid oscillation of output laser. It is difficult to get stable gain-switched pulse with high pump power and long pump pulse duration. Generally, fast gain-switching is adopted to achieve stable output pulse and a short pump pulse is needed for fast gain-switching. In the next step, we attempt to explore whether the stable output laser pulse can be generated in gain-switched fiber laser with both high pump power and long pump duration.

\section{Numerical analysis of the bias pumped gain-switched fiber laser}

\begin{figure*}[htbp]
\begin{minipage}{0.5\linewidth}
\centerline{\includegraphics[width=7cm,height=5cm]{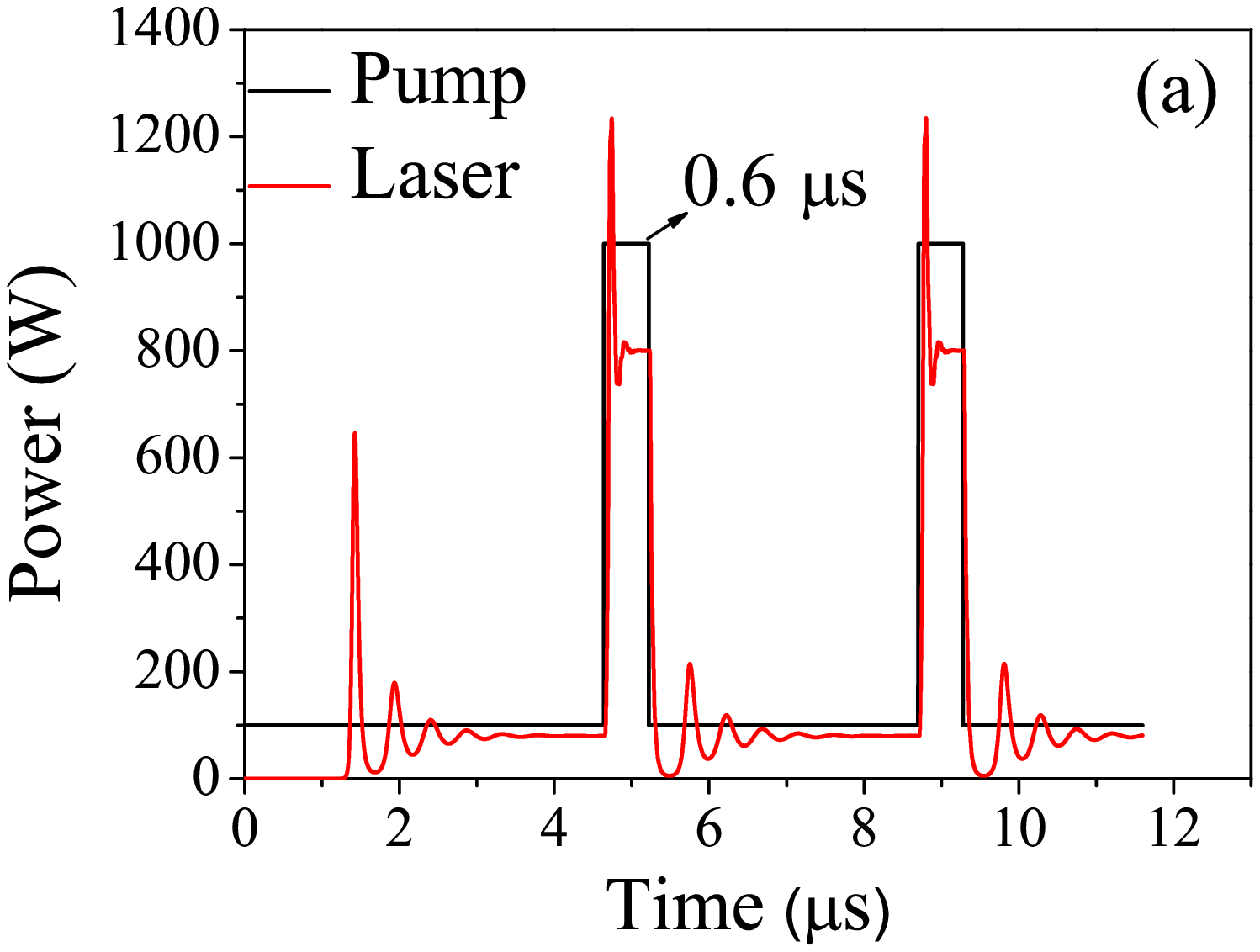}}
\centerline{ }
\end{minipage}
\hspace{-0.35in}
\begin{minipage}{0.5\linewidth}
\centerline{\includegraphics[width=7cm,height=5cm]{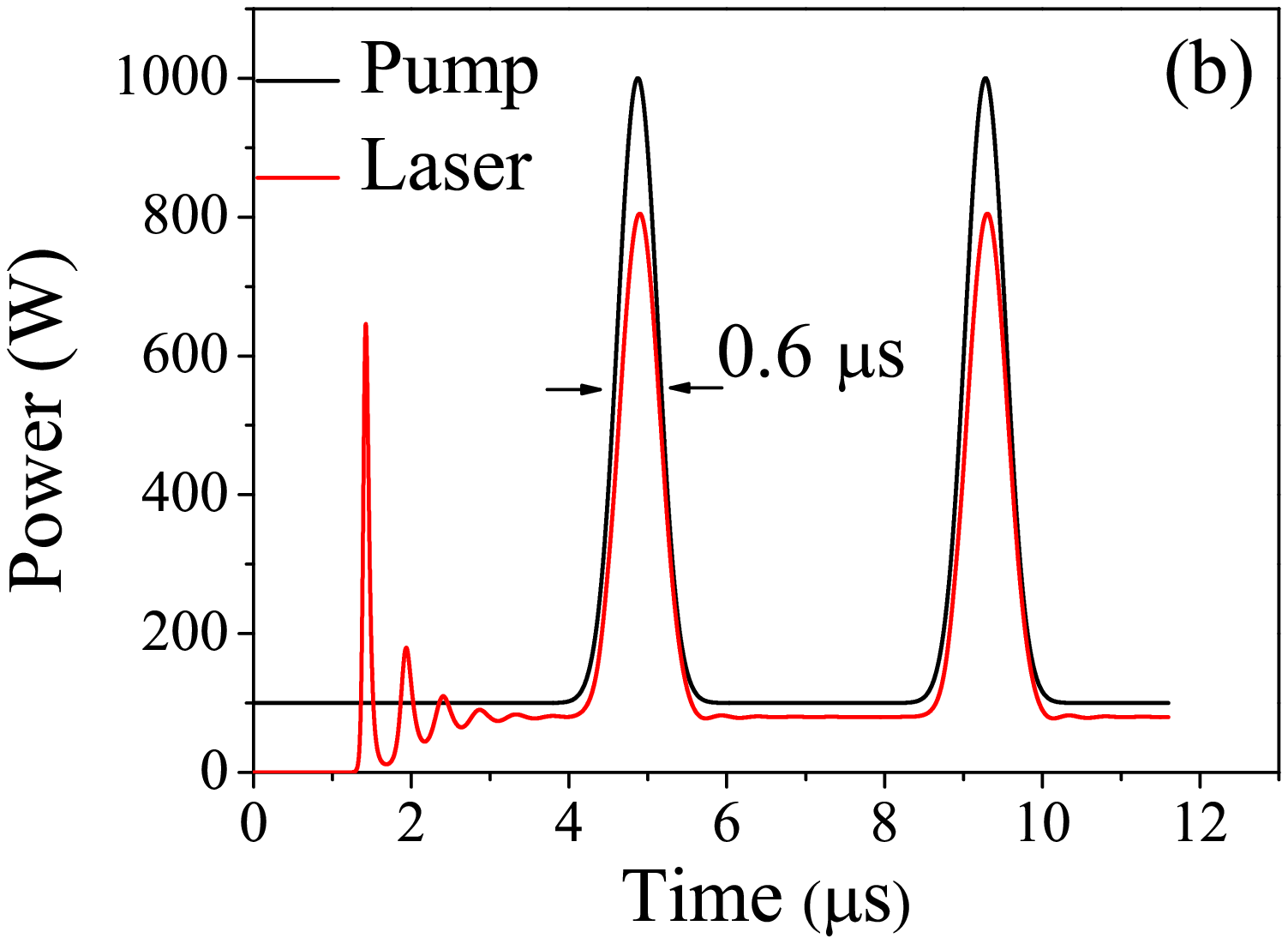}}
\centerline{ }
\end{minipage}
\vfill
\begin{minipage}{0.5\linewidth}
\centerline{\includegraphics[width=7cm,height=5cm]{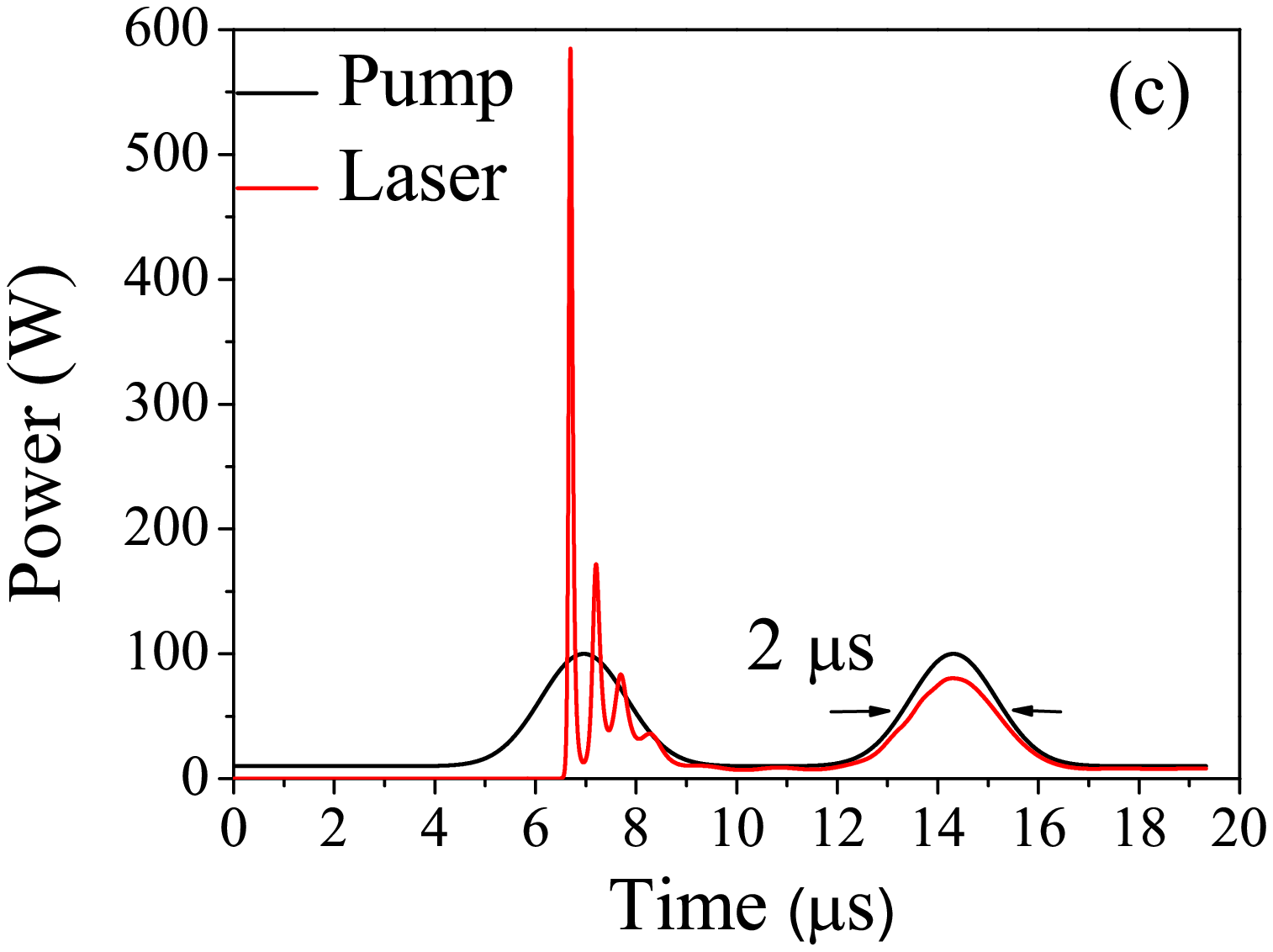}}
\centerline{}
\end{minipage}
\hspace{-0.35in}
\begin{minipage}{0.5\linewidth}
\centerline{\includegraphics[width=7cm,height=5cm]{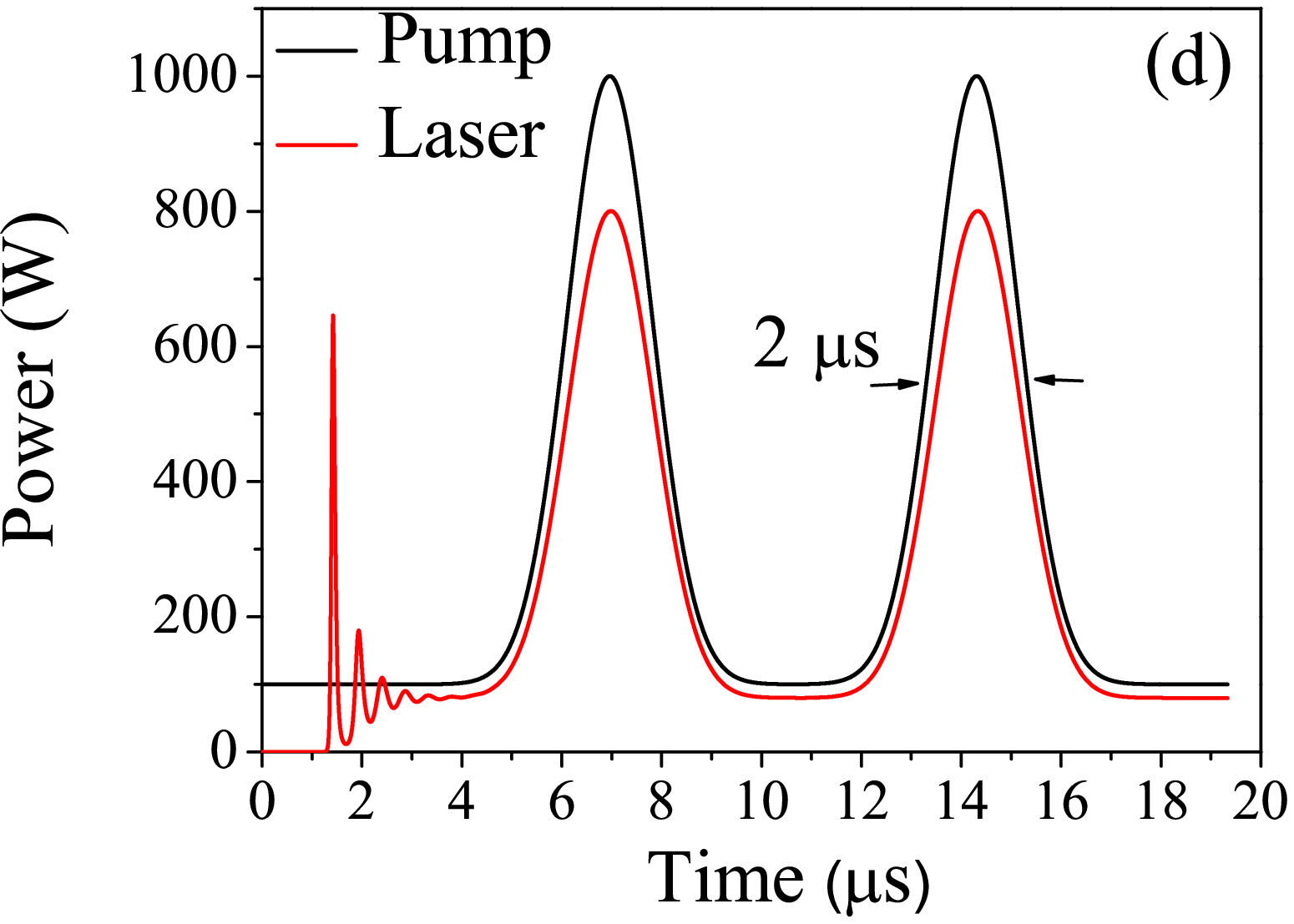}}
\centerline{ }
\end{minipage}
\caption{With 10 $\%$ pump power bias, temporal chracteristics of the outputs under pump peak power of 1000 W and duration of 0.6 $\mu$s (a) with rectangular pump pulse shape; (b) with gaussion pump pulse shape. With 10 $\%$ pump power bias, temporal chracteristics of the outputs under gaussion pump pulse shape and duration of 2 $\mu$s (c) with pump power of 100 W; (d) with pump power of 1000 W.}
\label{ss}
\end{figure*}

In conventional gain-switched fiber laser, pump power becomes zero when switching off the pump and the laser power also decays to zero. We also simulate the gain-switched fiber laser with pump power bias and the simulation results are shown in figure \ref{ss}. 

With a certain pump power bias, the emission laser power decays to a CW state instead of zero between two adjacent gain-switched output pulses. Compared to figure \ref{sr}(a) the spike caused by increasing pump power is greatly supressed with bias pumping as shown in figure \ref{ss}(a). However, the spikes are not regulated completely with rectangular pump profile. The sudden changing of pump is responsible for laser oscillation. We attribute laser oscillation in the front edge of output pulse to the steep profile of pump. Instead of using rectangular profile pump pulse, smooth pulses (gaussion pulse shape) are chosen as pump pulses and the simulation results are shown in figure \ref{ss}(b), (c) and (d). Compared to figure \ref{sr}(b) spikes caused by increasing pump peak power are regulated completely with bias pumping under gaussion pump profile as shown in figure \ref{ss}(b). Compared to figure \ref{sr}(d) spikes caused by long pump duration are regulated completely with bias pumping under gaussion pump profile as shown in figure \ref{ss}(c). Figure \ref{ss}(d) indicates bias pumping with smooth gaussion pump pulse can regulate chaotic relaxation spikes caused by both high power pumping and long pump duration. Therefore, bias pumping combined with smooth shape of pump pulse plays an important role in avoiding chrotic spikes of gain-switched pulse.
\begin{figure*}[htbp]
\centerline{
\includegraphics[width=7cm,height=6cm]{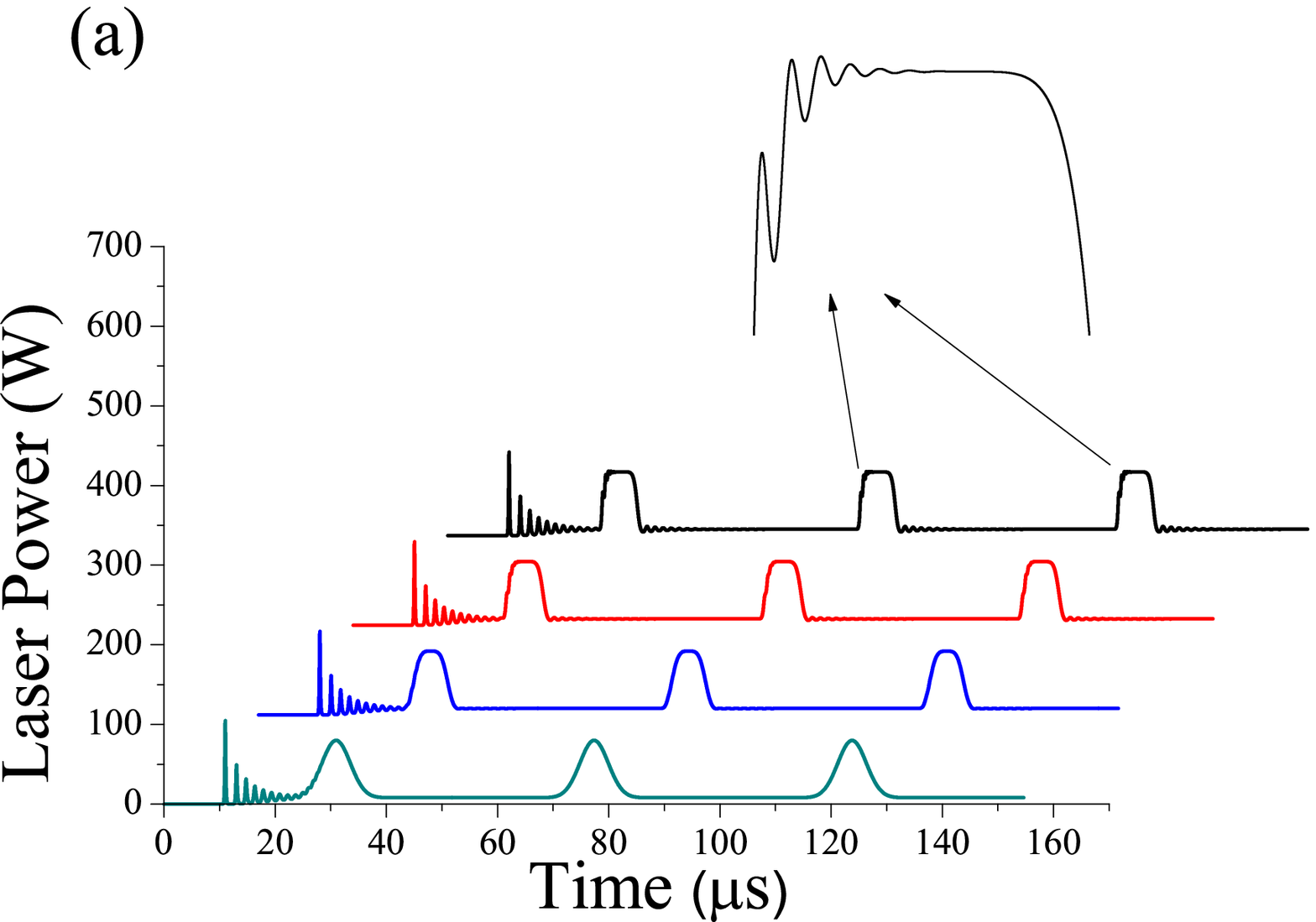}
\includegraphics[width=7cm,height=6cm]{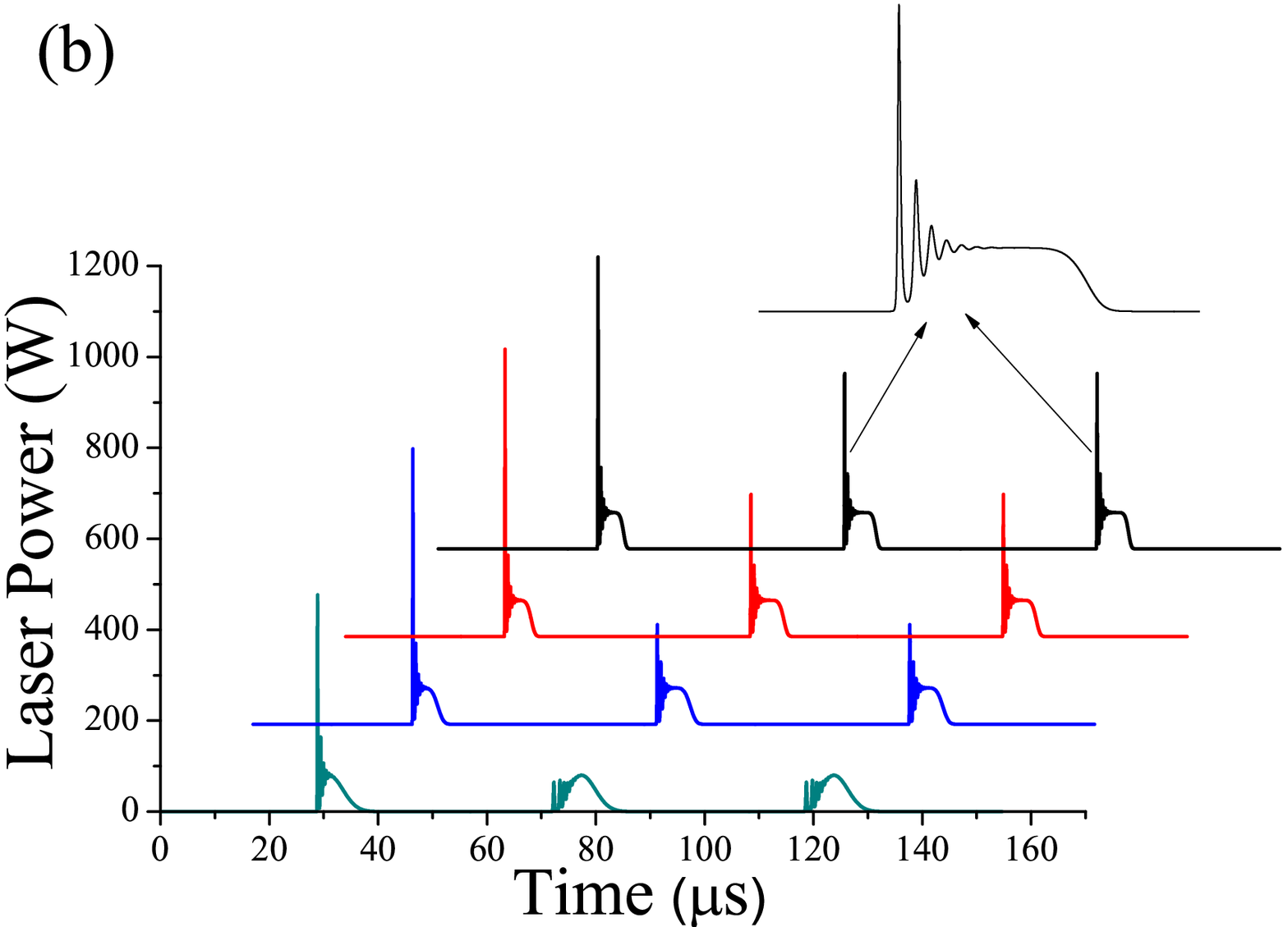}}
\caption{(a) Output gain-switched pulses with 10$\%$ pumping bias under different profiles of pump pulses. (b) Outpult gain-switched pulses without pumping bias under different profiles of pump pulses.}\label{st}
\end{figure*}

Another key factor that influences the temporal characteristics of gain-switched pulses is profile of pump pulse. Setting pump peak power of 100 W and duration of 6 $\mu$s, the temporal characteristics of output gain-switched pulses are simulated with different profiles of pump pulses and the results are shown in figure \ref{st} ((a) with pumping bias, (b) without pumping bias). In the simulation, the pump conditions are varied at the temporal profiles of pump pulses: gaussion profile, 4-order supergaussian profile, 6-order supergaussian profile and 8-order supergaussian profile. The corresponding output pulses are plotted with green line, blue line, red line and black line in figure \ref{st}. Except for the 8-order supergaussian profile case, the outputs are stable pulses with profiles almost identical with those of pump when adopting 10 $\%$ bias pumping in the other three cases (shown in figure \ref{st}(a)). In the 8-order supergaussian profile case, there are several spikes in the front edge of the output pulse. The higher the order, the steeper the edges of the profile are. High order supergaussion profile means quick changing of power in the edge of pump pulse, which results in relaxation oscillation of output laser. Without pump power bias, all of the output pulses consist of a series of relaxation spikes under four different pump profiles as shown in figure \ref{st}(b).  

The numerical results above indicate bias pumping is a potential technique to supress chaotic spiking. With bias pumping stable operation region of gain-switched fiber laser can be extended to long pump duration and high pump peak power, which can increase output pulse energy of stable gain-switched laser. what is more, under certain conditions the profiles of gain-switched stable pulse are almost identical with those of pump pulse, which may have potential application in simplifying the control unit of gain-switched fiber laser system. 

\section{Conclusion}

We have numerically demonstrated both long duration and high peak power of pump can lead to chaotic relaxation spike phenomenon in gain-switched fiber laser. In order to regulate those spikes, we proposed a bias pumped gain-switched fiber laser instead of shortening pump duration or decreasing pump peak power. The amplitude of chaotic relaxation spike in gain-switched fiber laser can be greatly supressed with bias pumping. Our simulation results show adopting bias pumping, under certain conditions, those chaotic relaxation spikes caused by long pump duration and high pump power can be completely eliminated. We also show smooth shape of pump pulse is required to avoid chaotic relaxation spikes in gain-switched fiber laser. In addition, the profile of output pulse from gain-switched fiber laser can keep the same shape as that of pump pulse with bias pumping, which may be beneficial to simplify the control unit. Compared to conventional gain-switched fiber laser, gain-switched fiber laser with bias pumping may have more applications due to its excellent output pulse chracteristics.  

We are very appreciate for Hong Po's valuable discussion.

\end{document}